\documentclass[aps,prl,twocolumn,groupedaddress,showpacs]{revtex4}
\usepackage[dvips]{graphicx}

\begin{document}
\title{Growing length and time scales in a suspension of athermal particles}
\author{Takahiro Hatano}
\affiliation{Earthquake Research Institute, University of Tokyo, 
1-1-1 Yayoi, Bunkyo, Tokyo 113-0032, Japan}
\date{\today}
\begin{abstract}
We simulate a relaxation process of non-brownian particles in a sheared viscous medium; 
the small shear strain is initially applied to a system, which then undergoes relaxation.
The relaxation time and the correlation length are estimated as functions of density, 
which algebraically diverge at the jamming density.
This implies that the relaxation time can be scaled by the correlation length 
using the dynamic critical exponent, which is estimated as $4.6(2)$.
It is also found that shear stress undergoes power-law decay at the jamming density, 
which is reminiscent of critical slowing down.
\end{abstract}
\maketitle

In general (thermal or athermal) particulate systems at high density, 
structural rearrangement of the constituent particles is difficult due to the exclusion volume effect
so that the structural relaxation time and the viscosity drastically increase.
In particular, zero temperature systems such as granular materials or emulsions 
acquire the elastic moduli above a certain density \cite{ohern1,ohern2}.
This rigidity transition, which is also referred to as the jamming transition, 
is accompanied by some power-law behaviors that are characteristic to critical phenomena: 
e.g. the growing correlation length in terms of spatially heterogeneous diffusion 
\cite{dauchot,durian1,durian2,durian3,lechenault}.
Note that, indeed prior to those in athermal systems, such dynamical heterogeneities 
are widely observed in thermal systems such as supercooled liquids and dense colloids 
\cite{hurley,muranaka,kob,yamamoto}.
This suggests the potential relation between the glass and jamming transitions, 
although still controversial \cite{liu,parisi,mari,berthier}.

Aside from the connection with glass transitions, however, 
the nature of the jamming transition itself is still not clear.
Provided that jamming is a critical phenomenon, the critical exponents that describe 
the divergence of the length and the time scales play an essential role 
in clarifying the underlying mechanism that yields the dynamical heterogeneity 
and classifying the jamming transition into a universality class (if any).
There are several attempts to estimate such exponents, in particular 
that for the correlation length; $\xi \sim |\phi-\phi_J|^{-\nu}$, 
where $\phi$ denotes the density and $\phi_J$ is the critical density.
Finite-size scaling reveals that $\nu\simeq0.7$ for both two and three dimensional systems, 
although the correlation length is not explicitly analyzed \cite{ohern2}.
Later, the correlation length is defined using the concept of the dynamical heterogeneity 
so that the exponent $\nu$ is estimated \cite{durian2,durian3,lechenault,drocco,olsson}.
Note that, however, the above attempts involve two dimensional systems 
and there is no direct estimation of the exponent $\nu$ in three dimensional systems.
Recalling that the dimensionality plays an essential role in conventional critical phenomena, 
one must investigate the three dimensional case.

Another important quantity that characterizes a critical phenomenon is the characteristic time.
Wyart et al. derive the characteristic frequency of jammed systems using the normal mode analysis.
They obtain $\omega_c\sim (\phi-\phi_J)^{\zeta}$, where $\zeta=0.5$ \cite{wyart}.
Interestingly, their discussion does not depend on the dimensionality of the system.
However, the characteristic time in unjammed systems is still not estimated.
Along the line of thought, in this paper, we investigate the behaviors of the relaxation time 
and the correlation length in a three dimensional unjammed system at zero temperature.
It is found that the relaxation time $\tau$ and the correlation length $\xi$ increases 
obeying power laws with respect to the density; 
$\tau\sim(\phi_J-\phi)^{-3.3}$ and $\xi\sim (\phi_J-\phi)^{-0.7}$.
These results lead to $z\simeq4.6$, which coincides with that of a Lennard-Jones glass 
\cite{berthier2}.

We consider macroscopic particles in a viscous medium 
so that temperature does not play any role.
We neglect hydrodynamic and electric interactions 
because we are interested in the nature of the jamming critical point.
The particles are monodisperse, the diameter of which is denoted by $d$.
Note that the particles are elastic so that the interaction between 
particles $i$ and $j$ is described by linear repulsive force; i.e., 
$f_{ij}=kh_{ij}{\bf n}_{ij}$, where $k$ denotes the elastic constant, 
${\bf n}_{ij}={\bf r}_{ij}/|{\bf r}_{ij}|$,  and $h_{ij}=d-|{\bf r}_{ij}|$ denoting the overlap length.
(Note that $h_{ij}=0$ if $d<|{\bf r}_{ij}|$).
We consider only unjammed systems, the volume density of which is 
less than the jamming density; $\phi<\phi_J\simeq0.639$ \cite{ohern2}.
Unless otherwise indicated, the system contains $6.4\times 10^4$ particles.

A relaxation process is realized in such a way that an equilibrated system is perturbed 
at $t=0$ and then undergoes time evolution.
To prepare an initial equilibrated system under periodic boundary conditions, 
we adopt the conjugate gradient method, by which randomly-distributed particles 
relax to a zero energy state (no overlaps between particles).
Then we perturb this equilibrated system by applying pure shear with respect to the $(y,z)$ plane; 
i.e., each particle is instantaneously displaced by the following Affine deformation.
\begin{eqnarray}
\label{affine1}
y'_i &=& y_i + \epsilon z_i \\ 
\label{affine2}
z'_i &=& z_i + \epsilon y_i,
\end{eqnarray}
where $\epsilon$ denotes the shear strain. At the same time, periodic boundary conditions 
are slightly modified in order to adapt the shear strain; 
the adjacent imaginary cells with respect to the $y$ and $z$ directions 
are displaced by $\epsilon L$ \cite{LE},  while an ordinary periodic boundary condition 
is used along the $x$ direction.
Due to the shear strain, overlaps between particles appear so that the system acquires 
nonzero elastic energy at $t=0$.
Then the system begins to relax and eventually reaches a new stable state of zero energy.
In order to mimic the dynamics of macroscopic particles in a highly viscous medium, 
we adopt overdamp dynamics; i.e., $\gamma\dot{{\bf r}}_{i} = \sum_{j} {\bf f}_{ij}$.
Throughout this study, we adopt the units in which the mass, the diameter, 
and the mobility $\gamma$ are unity.
The elastic constant $k$ is set to be $\gamma^2 /m$.
This procedure may be realizable in experiments using macroscopic particles 
in a viscous medium \cite{pine}, where the shear strain is applied via the viscous medium 
and the gravity can be canceled by density-matching.

\begin{figure}
\includegraphics[scale=0.4]{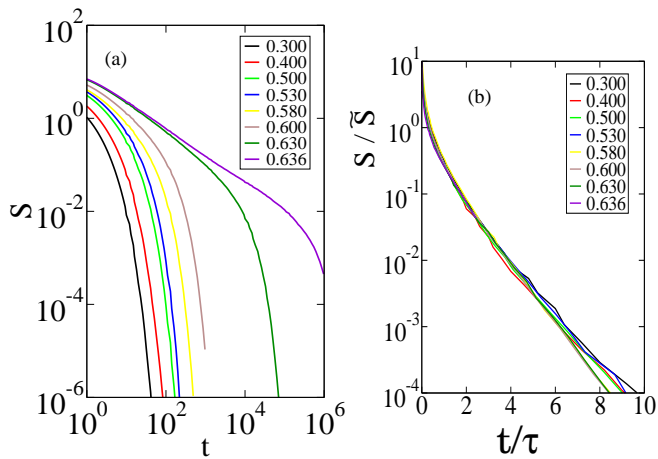}
\caption{\label{stress} 
(a) The relaxation of shear stress at various densities (see the legends).
Lower curves correspond to lower densities.
The relaxation time grows as the density increases towards the critical density.
(b) Rescaled shear stress $S / \tilde{S}$ as a function of rescaled time, $t/\tau(\phi)$, 
where $\tau(\phi)$ is the characteristic time that depends on the density, 
and $\tilde{S}$ is approximately $0.5S_0/\sqrt{\tau}$.}
\end{figure}
First, we discuss the relaxation of macroscopic quantities, in particular shear stress, 
which is defined via the virial \cite{zubarev}.
Because the temporal behavior of other macroscopic quantities such as pressure and energy 
are essentially the same, we focus shear stress here.
The relaxation of shear stress at each density is shown in FIG. \ref{stress} (a), 
where the initial strain $\epsilon$ is $1.0\times 10^{-3}$. 
We also test $\epsilon=0.01$ and $\epsilon=0.05$ to find that 
the result does not qualitatively depends on the initial strain.
It is important to remark that several samples at each density show quantitatively 
the same relaxation behavior; 
less than $\pm10$ \% in the characteristic time defined below.

Then we define the characteristic time from the relaxation behaviors shown in FIG. \ref{stress} (a).
These temporal behaviors indeed collapse by rescaling the time with 
the characteristic time at each density, $\tau(\phi)$, as shown in Fig. \ref{stress} (b).
Note that shear stress is also rescaled by $\tilde{S}\sim0.5S_0/\sqrt{\tau}$, 
which indicates that the temporal behavior of shear stress is described by  
\begin{equation}
\label{Srelax}
S(t) \sim S_0 t^{-\alpha} e^{-t/\tau},
\end{equation}
with $\alpha\simeq0.5$.
\begin{figure}
\includegraphics[scale=0.4]{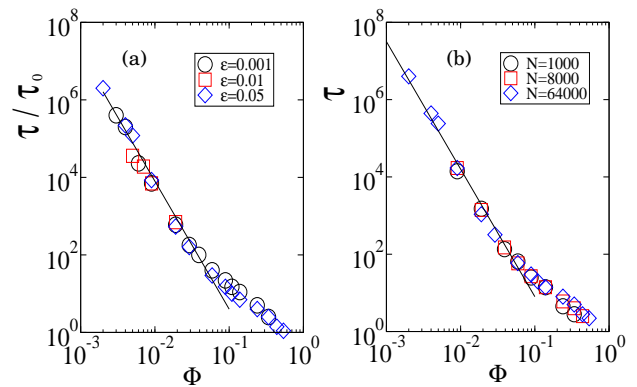}
\caption{\label{tau_fig}
The relaxation time as a function of density.
Solid line in each panel is proportional to $\tau\sim\Phi^{-3.3}$, 
where $\Phi=\phi_J-\phi$ and $\phi_J=0.639$.
(a) The relaxation time normalized by $\tau_0$: $\tau_0=2$ for $\epsilon=0.05$ , 
$\tau_0=1$ otherwise. The power-law divergence and the exponent 
do not depend on the initial strain.
(b) There is no system-size dependence of the relaxation time. The initial strain $\epsilon=0.05$. }
\end{figure}
By fitting the shear stress relaxation with Eq. (\ref{Srelax}), we obtain $\alpha=0.55(5)$.
The relaxation time $\tau(\phi)$ drastically increases at higher densities as shown 
in FIG. \ref{tau_fig} (a), which indicates the power-law divergence of the relaxation time 
at the jamming density.
\begin{equation}
\label{tau}
\tau \sim \tau_0(\phi_J-\phi)^{-\zeta},
\end{equation}
where $\tau_0$ denotes the time constant that does not depend on the density 
and the exponent is estimated as $\zeta=3.3(1)$.
However, note that this exponent is valid only in higher density region, 
$\phi \ge 0.60$.
There seems to be a crossover to the different behavior in lower density region, $\phi\le0.60$, 
where $\zeta=1.5(1)$ if we assume Eq. (\ref{tau}).
Note that this is not due to a finite-size effect as illustrated in FIG. \ref{tau_fig} (b), 
which indicates the relaxation time for three different systems: 
$N=1000, 8000, 64000$.
This is indeed due to the qualitative change in the particle dynamics, 
because the correlation function discussed below, Eq. (\ref{Gr}), 
cannot detect any cooperative motion in the lower density region, $\phi<0.60$.
However, as we focus on the critical nature of the jamming transition here, 
we do not further discuss this crossover.

\begin{figure}
\includegraphics[scale=0.35]{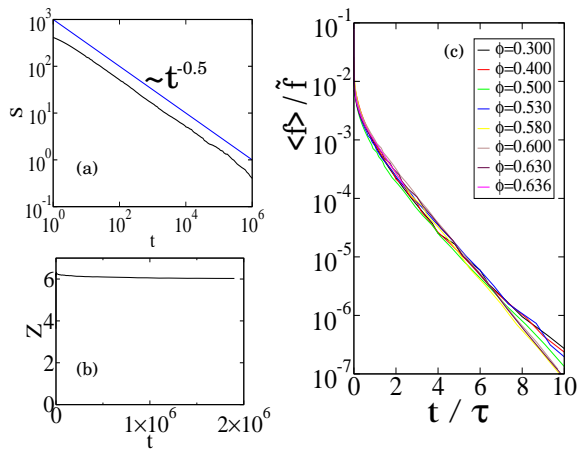}
\caption{\label{SZh}
(a) The relaxation of shear stress for $\phi=0.637$ and $\epsilon=0.05$.
The solid line is proportional to $t^{-0.5}$.
(b) The relaxation of the average coordination number for $\phi=0.637$ and $\epsilon=0.05$.
(c) The relaxation of the average magnitude of interparticle force, $\langle f\rangle$.
Time is rescaled by $\tau(\phi)$ shown in FIG. \ref{tau_fig} (a).
Note that $\langle f\rangle$ is also rescaled by ${\tilde f}\sim \tau^{0.3}$.
The initial strain is $0.001$.}
\end{figure}

From Eqs. (\ref{stress}) and (\ref{tau}), it is expected that the relaxation 
of shear stress is described by a simple power law at the jamming density, 
$S(t)\sim S_0 t^{-\alpha}$.
Figure \ref{SZh} (a) shows the stress relaxation at $\phi=0.637$, 
which obeys a power-law decay for a considerable duration (up to six orders of magnitude).
We remark that this power-law relaxation, which is quite similar to critical slowing down, 
is consistent with a theory in which shear stress is an order parameter 
that undergoes marginal stability at the jamming transition point \cite{otsuki,hatano}.

The slow relaxation is also apparent in view of the particle dynamics.
A quantity that involves the stresses and energy is the magnitude of interparticle force.
We thus define the average magnitude of interparticle force as 
\begin{equation}
\label{f}
\langle f \rangle = \frac{2}{NZ}\sum_{i>j} |{\bf f}_{ij}|, 
\end{equation}
where $Z$ is the coordination number.
We remark that $Z$ is almost constant during the relaxation process as shown in FIG. \ref{SZh} (b),
while the average magnitude of force relaxes as 
\begin{equation}
\label{frelax}
\langle f \rangle \propto t^{-\beta} e^{-t/\tau},
\end{equation}
where $\beta\simeq0.3$.
The relaxation profile of $\langle f \rangle $ is shown in FIG. \ref{SZh} (c), 
where the time is rescaled by the relaxation time $\tau$ defined by Eq. (\ref{Srelax}).
Furthermore, we observe that energy and pressure relax in essentially the same manner; 
i.e., $P\propto t^{-\beta} e^{-t/\tau}$ and $E\propto t^{-2\beta} e^{-2t/\tau}$.
This is indeed trivial because $P\sim 2(NZ)^{-1}\langle f \rangle d/V$ and 
$E\sim 2(NZ)^{-1}\langle f^2\rangle /kV$, where $d$ and $k$ denote the diameter 
and the stiffness of the particles, respectively.
We also remark that $f$ obeys an exponential distribution so that 
$\langle f^2\rangle\sim \langle f\rangle^2$.

\begin{figure}
\includegraphics[scale=0.4]{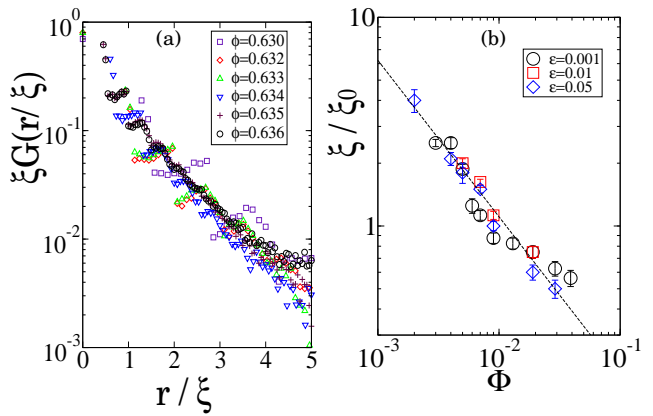}
\caption{\label{xi} 
(a) The collapse of the correlation function defined by Eq. (\ref{Gr}).
Note that the length is normalized by the correlation length, $\xi$.
(b) The correlation length as a function of $\Phi=\phi_J-\phi$, where $\phi_J=0.639$. 
The dashed line is proportional to $\Phi^{-0.75}$.
Note that the correlation length is normalized by $\xi_0$, 
which depends on the initial strain $\epsilon$.
Here $\xi_0$ is set to be unity for $\xi_0=0.05$, while $\xi_0=0.5$ for $\epsilon=0.001$ 
and $\xi_0=0.8$ for $\epsilon=0.01$.}
\end{figure}

We then investigate the spatial correlation of the particle motion during the relaxation process.
To this end, we compare the initial and the final configurations to define 
the displacement vector of each particle; 
$\Delta{\bf x}_i = {\bf x}_i^{(1)} - {\bf x}_i^{(0)} - {\bf A}({\bf x}_i^{(0)})$, 
where ${\bf x}_i^{(0)}$ and ${\bf x}_i^{(1)}$ are the initial and the final positions of particle $i$, 
and ${\bf A}({\bf r})$ is the Affine deformation vector at position ${\bf r}$, 
represented by Eqs. (\ref{affine1}) and (\ref{affine2}).
Using the displacement vectors $\Delta{\bf x}_i $, we define the following correlation function 
\cite{muranaka,doliwa}.
\begin{equation}
\label{Gr}
G(r) = \frac{\sum_{i>j}\Delta{\bf x}_i\cdot\Delta{\bf x}_j\delta(r-|{\bf x}_i^{(0)}-{\bf x}_j^{(0)}|)}
{\sum_{i>j}\delta(r-|{\bf x}_i^{(0)}-{\bf x}_j^{(0)}|)}.
\end{equation}
This quantifies the extent of chainlike cooperative motion, which is universally observed 
in dense particulate systems; i.e., supercooled liquids, colloids, and grains 
\cite{durian3,donati,marcus,ediger}.
Figure \ref{xi} (a) shows the collapse of the correlation function, 
where $\xi G(r/\xi)$ is plotted as a function of $r/\xi$.
This indicates that $\xi$ is the correlation length 
and the correlation function $G(r)$ is approximately exponential.
As is shown in FIG. \ref{xi} (b), this correlation length $\xi$ increases 
as the density approaches the jamming density; 
$\xi\sim\xi_0(\phi_J-\phi)^{-\nu}$, where $\nu\simeq0.7$.
This value is consistent with the shift exponent for the static rigidity transition 
in three dimensions \cite{ohern2}.
Also, within the present numerical accuracy, it is indistinguishable from the exponent 
obtained from the dynamical heterogeneity in two dimensional systems 
\cite{dauchot,durian1,durian2,durian3,lechenault,drocco} 

\begin{figure}
\includegraphics[scale=0.25]{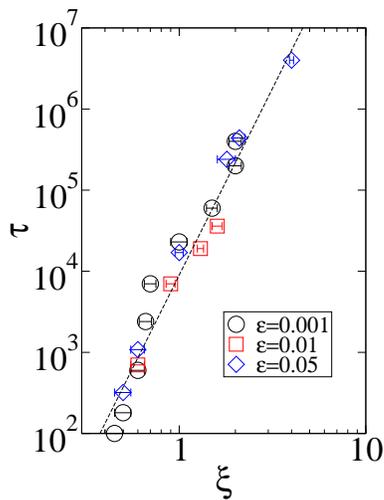}
\caption{\label{xi-tau} 
The correlation length (normalized by $\xi_0$) dependence of the relaxation time 
(normalized by $\tau_0$). Data are taken from FIGS. \ref{tau_fig} and \ref{xi}.
The dashed line is proportional to $\xi^{4.6}$.}
\end{figure}

From the present numerical results we can estimate the dynamic critical exponent, $z=\zeta/\nu$.
We replot the correlation length and the relaxation time in FIG. \ref{xi-tau}, 
where we can estimate $z$ as $4.6(2)$.
This value is quite different from that obtained in two-dimensional air-fluidized beads, 
where $z\simeq 1.4$ \cite{durian2}.
Although the difference may be due to the dimensionality, 
we do not have any definite explanation on this difference at this point.
The structural relaxation process may strongly depend on the specific dynamics, 
particularly an agitation method by which energy is injected to the system.
Nevertheless, it is noteworthy that almost the same dynamic critical exponent 
($z\simeq4.55\pm0.2$) is found in a binary Lennard-Jones (LJ) supercooled liquid, 
where the correlation length and the characteristic time for the structural relaxation 
diverges towards zero temperature \cite{berthier2}.
This result, together with a recent study \cite{berthier}, implies that 
a binary LJ glass is dominated by a zero temperature critical point, 
which may have close relation to the jamming transition.
Although the correspondence between an LJ glass and the present system is not apparent, 
this coincidence may provide another evidence for the close relation 
between the glass and jamming transitions.

To summarize, we show power-law divergence of the relaxation time and the correlation length 
as the density approaches the jamming density from below; $\tau/\tau_0\sim(\phi_J-\phi)^{-\zeta}\sim(\xi/\xi_0)^{z}$, 
where $\zeta=3.3(1)$ and $z=4.6(2)$.
The correlation length is defined in terms of stringlike cooperative motion, 
which is widely observed in dense particulate systems.
It is also found that the stress relaxation near the jamming density obeys critical slowing down.
Although the result is restricted to a three dimensional unjammed system, 
we remark that the length and time scales in jammed systems (where $\phi > \phi_J$) 
and the effect of dimensionality is currently investigated by using the present relaxation method.

The author is grateful to Shin-ichi Sasa, Hiroki Ohta, and Michio Otsuki for helpful discussions.

\end{document}